\documentstyle[aps,prl,twocolumn,tighten,epsbox]{revtex}

\begin{document}

\title{Resistance Jumps and Hysteresis in Ferromagnetic Wires}

\author{Tohru Koma$^1$$^*$ and Masanori Yamanaka$^2$$^{\dagger}$}

\address{$^1$Department of Physics, Gakushuin University, 
Mejiro 1-5-1, Toshima-ku, Tokyo 171-8588, Japan}

\address{$^2$Department of Applied Physics, Science University of Tokyo, 
Kagurasaka 1-3, Shinjuku-ku, Tokyo 162-8601, Japan}

\maketitle

\begin{abstract}
A phenomenological theory based on the principle of minimum heat 
generation is presented for the recently discovered exotic electric 
transport in ferromagnetic wires with a domain wall. 
We provide an unified explanation of the negative 
and positive jumps and of the hysteresis loops observed 
in the magnetoresistance experiments. 
Our argument is based on a microscopic calculation which shows that 
the presence of a domain wall sometimes enhances the transmission 
probability through an impurity potential, as well as the 
phenomenological principle of minimum heat generation. 
\end{abstract}


The Ohm's law, which states that the measured voltage 
is proportional to the current flow, 
has been observed in many conducting materials. 
It is usually regarded that Ohmic transport can be explained by 
means of a suitable linear response theory. 
Non-Ohmic transports, where the voltage has a nonlinear dependence 
on the current, are also observed in many conductors. 
Unlike Ohmic transport, non-Ohmic transport usually results from 
complicated nonlinear responses of a microscopic state 
which takes place when a finite current flows through the material. 

Recently, in a series of experiments in metallic wires 
containing ferromagnetic domain walls \cite{HG,Otani,OtherNegative,Parkin}, 
exotic non-Ohmic transport phenomena were observed. 
The transport properties exhibit a delicate dependence on the external 
magnetic field which controls the motion of the domain walls. 
Most remarkably, the non-Ohmic resistance 
(defined simply as the ratio between the voltage and the current) 
shows an abrupt negative jump and a hysteresis loop 
when the magnetic field is varied. 
It is believed that these peculiar behaviors are caused by the motion 
of the ferromagnetic domain wall(s). 

Although some microscopic mechanisms were proposed 
\cite{Parkin,TF,OtherTheory} in order to explain these phenomena, 
no satisfactory unified picture exists at present for the whole phenomena, 
and there still remain many issues (e.g., the magnitudes of the resistance 
jumps, and the shape of the hysteresis loop) to be explained. 
Clearly the difficulty comes from the nonlinearity and the complexity. 
The standard linear response theory is obviously useless 
in dealing with non-Ohmic transport. 
We also expect that various effects such as 
the crystal anisotropy, the Lorentz force, and 
the scattering by various scatterers (e.g., impurities, 
domain walls, localized spins, and the surface of the sample) 
are entangled with each other in a complex manner thus producing 
the observed transport properties. 
It seems unlikely to us that one would be able to treat these effects 
separately. A phenomenological approach which deals with overall properties 
of the system as a whole seems most suitable in studying 
such complex nonlinear phenomena. 

In the present Letter we propose a phenomenological explanation for 
the non-Ohmic nature, the negative and positive jumps, 
and the hysteresis loops, in the magnetoresistance experiments for 
the ferromagnetic wires. 
Our basic strategy is to apply the principle of minimum heat generation. 
The principle states that, when a given amount of electric current 
goes through a sample, the local currents distribute themselves so as 
to make the total heat generation rate as small as possible. 
(See, e.g., Chap.~19 of Ref.~\cite{FLN}.) 
As is well known, when the resistance (or the resistance distribution) 
of the sample is independent of the current, the principle leads to Ohmic 
(or the Kirchhoff's) law. 
When the resistance depends on the current, 
then the principle may lead to a nonlinear voltage-current relation. 

\begin{figure}
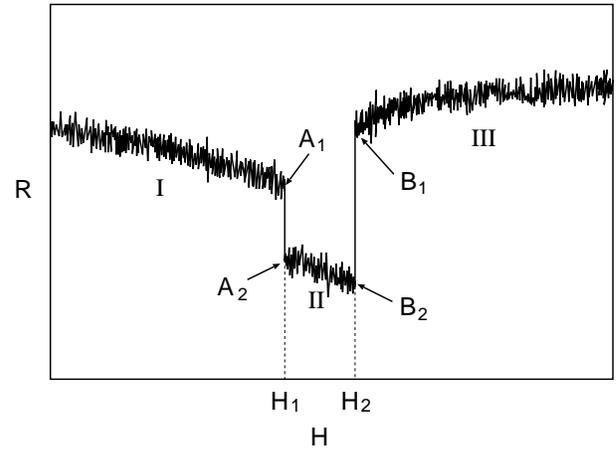

\epsfile{file=f1.eps,width=80mm}
\smallskip\smallskip
\caption{A typical resistance curve $R$ as a function of an external 
magnetic field $H$. The curve consists of three stages 
and two resistance jumps.}
\label{Rcurve}
\end{figure}
Figure~\ref{Rcurve} schematically shows 
a typical experimental result \cite{Otani} 
of the magnetoresistance $R$ of a ferromagnetic wire 
under a uniform magnetic field $H$ parallel to the current. 
In the initial state, the wire is set in a strong magnetic field 
(which is taken to be negative for the following convenience) 
and all the domain walls are swept out of the wire. 
The magnetic field then is increased from 
the initial negative value and swept towards positive values, 
in order to create and control a single domain wall in the wire. 
In Stage~I in Fig.~\ref{Rcurve}, the resistance $R$ 
decreases with fluctuations as the magnetic field $H$ is increased. 
It has been confirmed experimentally that there is no domain wall in the wire 
at this stage. 
At the magnetic field $H_1$, the resistance $R$ abruptly jumps from 
the point A$_1$ to A$_2$, and this negative jump of the resistance $R$ 
is accompanied with an abrupt appearance of a single ferromagnetic domain wall 
in the wire. In Stage~II, the resistance $R$ decreases with fluctuations 
as the magnetic field $H$ is increased, and it was observed 
that the single domain wall remains in the wire. 
At the magnetic field $H_2$, the resistance $R$ abruptly jumps from 
the point B$_2$ to B$_1$, and this positive jump of the resistance $R$
is accompanied with an abrupt disappearance of the domain wall. 
After this resistance jump, the magnetization of the wire is fairly saturated 
in the positive direction parallel to the magnetic field. 
In the final Stage~III, the resistance $R$ increases with fluctuations 
as the magnetic field $H$ is increased, and no domain wall is observed 
in the wire. This resistance curve $R$ is not 
reversible for the reversal process where the magnetic field $H$ is gradually 
decreased. Hong and Giordano \cite{HG} pointed out 
that this hysteretic behavior of the resistance $R$ is closely related to 
the hysteresis of the magnetization process 
and to the existence of a domain wall in the ferromagnet. 

In order to understand this behavior, 
consider a metallic wire with impurities and a ferromagnetic domain wall. 
Though the impurities potential is fixed, 
there remain many degrees of freedom which affect 
the transport properties, and it will be shown that 
they indeed lead to non-Ohmic characters. 
As mentioned above, these degrees of freedom include 
the position of a domain wall and the structure of magnetic domains. 
According to the principle of minimum heat generation, 
they distribute themselves in the wire 
so as to minimize the total heat generation rate. 

We begin with a simple situation where 
there is only one domain wall in the wire. 
When there is no trapping potential of impurities, 
the domain wall freely moves in the wire. 
When there is a fixed impurities potential, 
which position does the domain wall favor? 
If the domain wall which entered the wire 
is abruptly trapped at a special position that 
minimizes the scattering amplitude of the conduction 
electrons from the impurities potential, 
then a negative resistance jump as in Fig.~\ref{Rcurve} 
is expected to take place. 
In order to justify this picture, we use the effective 
one-dimensional Schr\"odinger equation \cite{CF} with 
the single domain wall with the center at $x=x_0$, 
\begin{eqnarray}
& &\left[-\frac{\hbar^2}{2m}\frac{d^2}{dx^2}
+\frac{J \Delta}{4}\tanh\left(\frac{x-x_0}{\lambda}\right)\sigma_x
+V_{\rm imp}(x)\right]
\psi(x)\nonumber\\
&=&\frac{J}{4}{\rm sech}\left(\frac{x-x_0}{\lambda}\right)\sigma_z\psi(x)
+E\psi(x)
\label{eq:schrodingereq}
\end{eqnarray}
for the conduction electron with an energy $E$, the mass $m$, and 
with the wavefunction, 
$$
\psi(x)=\left(\matrix{\psi_\uparrow(x) \cr 
\psi_{\downarrow}(x)\cr}\right); \
\sigma_x=\left(\matrix{1 & 0\cr 0 & -1\cr}\right)\ \mbox{and}\ 
\sigma_z=\left(\matrix{0 & 1\cr 1 & 0\cr}\right). 
$$
The spin $s=\uparrow,\downarrow$ of the electron interacts 
with the ferromagnetic domain wall
through the effective potentials $J\Delta\tanh[(x-x_0)/\lambda]\sigma_x/4$ 
and $-J{\rm sech}[(x-x_0)/\lambda]\sigma_z/4$, where $J$ is 
the Hund coupling, $\Delta$ the anisotropy and $\lambda$ the width 
of the domain wall. 
The domain wall should be realized on the quantum ferromagnetic 
Heisenberg model, and the width $\lambda$ is determined by 
the exchange anisotropy of the Heisenberg model \cite{QDW}. 
See Ref.~\cite{YK} for details of the microscopic derivation of 
the effective Schr\"odinger equation (\ref{eq:schrodingereq}). 
We take the impurity potential 
$V_{\rm imp}(x)$ to be a rectangular barrier as 
\begin{eqnarray}
V_{\rm imp}(x) = \left\{\begin{array}{ll}
V_0, & \mbox{if}\ \vert x \vert \le \frac{w}{2}\\
0, & \mbox{otherwise}
\end{array}
\right.
\end{eqnarray}
with the width $w$ and the height $V_0>0$. 
As we shall see below, details of the impurity potential 
do not affect our phenomenological argument to 
understand the experimental results. 
We assume that the potential $V_{\rm imp}(x)$ is fixed (i.e., quenched). 
Although the position $x_0$ of the domain wall varies to 
realize the minimum heat generation, we assume that 
the position $x_0$ can be treated as a fixed parameter 
in Eq.~(\ref{eq:schrodingereq}) because 
the motion of the domain wall is much slower than that of 
conduction electrons. 

\begin{figure}
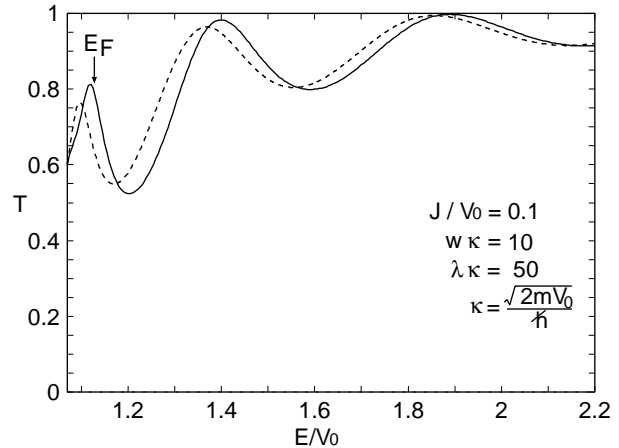

\epsfile{file=f2.eps,width=80mm}
\smallskip\smallskip
\caption{Transmission probabilities $T$ as a function of energy $E$ 
of the conduction electron through the rectangular potential 
without (dashed line) and with (solid line) the single domain wall 
sitting right above the potential.} 
\label{Tresult}
\end{figure}
We numerically solved the above Schr\"odinger equation 
(\ref{eq:schrodingereq}) for the case $x_0=0$ (i.e., the domain wall 
sits right above the impurity potential), 
and obtained the transmission probabilities 
$T=(T_\uparrow+T_\downarrow)/2$ for a single electron. Here $T_s$ is the 
transmission probability for the incident plane wavefunction 
with the polarized spin $s$, and for simplicity, we have assumed 
that the spin up and down electrons near the Fermi level $E_{\rm F}$ 
equally contribute to the transport. This assumption does not 
affect our conclusion below because $T_\downarrow$ exhibits a very similar
behavior to $T_\uparrow$, where we varied the energy $E$
in our numerical results. Figure~\ref{Tresult} shows 
the calculated transmission probabilities through
the rectangular impurity potential with and without the domain wall. 

In the case with the simple rectangle barrier and without domain walls, 
the transmission probability $T$ (dashed line in Fig.~\ref{Tresult}) 
oscillates between the minimum and the maximum 
values as the energy $E$ of the electron increases. 
When the domain wall is coupled to the impurity potential, 
the same quantity (solid line in Fig.~\ref{Tresult}) 
shows a similar oscillation but with a different phase. 
As a consequence, we find that there are some ranges of energy $E$ 
in which the transmission probability $T$ with the domain wall 
is strictly greater than that without a domain wall. 
In the generic case for the position of the domain wall, 
we assume that the distance $|x_0|$ between the domain wall 
and the impurity is bounded as $|x_0|\le {\bar x}$ with a positive 
constant ${\bar x}$, since actual impurities are almost homogeneously 
distributed in a wire. In other words, there exists at least one impurity 
within a region $[a,b]$ with $\vert b-a \vert=2{\bar x}$. 
Under this assumption, we obtain, from the results in Fig.~\ref{Tresult}, 
the following: 
At least for electrons with energy $E$ in one of the above ranges, 
there exists an optimal position for the domain wall 
at which the transmission probability $T$ is maximized and 
hence the heat generation is minimized \cite{MP}. 

Consequently we find from our numerical results for the position $x_0=0$ 
of the domain wall that the presence of a domain wall enhances 
the transmission probability through an impurity potential in the wire 
if the Fermi energy $E_{\rm F}$ of the conduction electrons is 
in the above energy range.
In a realistic situation, the potential for most impurities, however, 
is not of a simple rectangular form. 
But, since real impurities are randomly distributed in a wire, 
we expect that a potential consisting of many rectangular potentials 
having various widths and various heights yields the same effect as 
that of a realistic impurities potential. 
Furthermore, from the above result for a single rectangular potential, 
we find that the transmission probability through the wire is enhanced 
by a single domain wall trapped at a special rectangular potential 
that satisfies the above energy condition. 
Combining these observations with the principle of minimum heat generation, 
we conclude that a single domain wall entered in the wire 
is trapped at the special position that minimizes the scattering amplitude 
of the electron at the Fermi level $E_{\rm F}$. 
This explains the negative resistance jump from the point A$_1$ to A$_2$ 
in Fig.~\ref{Rcurve}. 
In this description, the magnitude $\Delta R$ of the resistance jump
is evaluated as $\Delta R\approx\Delta TR\lambda/L$ with the length $L$ 
of the wire and the increase $\Delta T$ of the transmission probability 
due to the domain wall. Substituting the experimental data \cite{Otani}, 
$\Delta R\approx 0.005 \ \Omega$, $R\approx 41\ \Omega$, $\lambda=20$ nm 
and $L=20\ \mu$m, into this relation, we have $\Delta T\approx 0.12$ 
which is consistent with our numerical result $\Delta T\approx 0.15$ 
which is read from Fig.~\ref{Tresult}. 

\begin{figure}
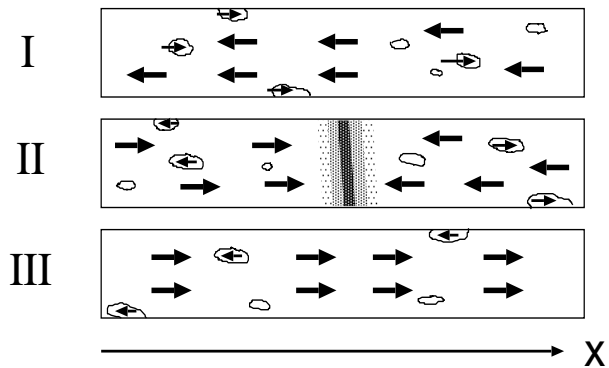

\epsfile{file=f3.eps,width=80mm,height=50mm}
\caption{Typical snapshots of the magnetization profiles in the three 
stages in the resistance curve $R$.} 
\label{ThreeStages}
\end{figure}

In order to explain Stage~I in Fig.~\ref{Rcurve}, 
consider first a simple situation that there is only one small island 
of localized up spins in the background sea of localized down spins. 
Note that, if there is no impurity in the wire, the position of 
the island is not determined by the energetics alone 
because of the translational invariance of the system. 
When many impurities exist in the wire, the position of the island is 
determined by the principle of minimum heat generation, 
so that the scattering of the electrons at the Fermi level 
$E_{\rm F}$ by the impurities potential 
is suppressed by the presence of the island. 
Here the mechanism of the suppression for the scattering 
is essentially the same as in the above case with a single domain wall. 
Having this result in mind, let us consider the situation 
in Stage~I in Fig.~\ref{Rcurve}. 
In this stage, there appear many islands of up spins 
as the external magnetic field $H$ is increased, as 
schematically shown in Fig.~\ref{ThreeStages}. 
The spatial configuration of these islands is also 
determined by the principle of minimum heat generation so that 
each island suppresses the scattering of the conduction electrons 
by an impurity, by combining itself with the impurity. 
In consequence, the resistance $R$ of the wire decreases 
as the number of the islands increases. 
Since the number of the islands increases as the magnetic field $H$ is 
increased, we recover the observed magnetic filed dependence of 
the resistance $R$. Moreover, since the total size of the islands is 
roughly proportional to the total magnetization of the wire, we see 
that, roughly speaking, the resistance $R$ in Stage~I 
must be a function of the total magnetization. 
This clearly explains why the resistance curve $R$ shows hysteretic 
behaviors as those found in the magnetization curve of the ferromagnet. 
For the same reason, we expect similar hysteretic behaviors in 
Stages~II and III. 

In the same way as in Stage~I, we can explain Stage~II. 
Namely the increase of the number of up spin islands 
in the sea of down spins decreases the resistance $R$, 
while a single domain wall trapped at a special position also 
suppresses the scattering of the electrons by the impurities potential. 

With the abrupt resistance jump from the point B$_2$ to B$_1$, 
the domain wall and many islands of up spins disappear from the wire. 
Since these degrees of freedom contributed in decreasing 
the resistance $R$, their disappearance leads to 
an abrupt positive resistance jump as in Fig.~\ref{Rcurve}. 
After this abrupt change, i.e., in Stage~III, the magnetization of the wire is 
nearly saturated. In other words, there sparsely exist down spin islands 
in the background sea of up spins, as schematically shown 
in Stage~III of Fig.~\ref{ThreeStages}. 
As the magnetic field $H$ is further increased, 
the down spin islands disappear or become smaller, and finally 
all of them disappear from the wire. 
Since these degrees of freedom also had an effect of decreasing 
the resistance $R$, their gradual disappearance leads to an increase of 
the resistance $R$ observed in Stage~III. 

Let us conclude by making three remarks. 
When a single domain wall enters in the wire 
at the magnetic field $H_1$, most of up spin islands are 
expected to be swept out of the left-hand side region of the domain wall. 
(See Stages~I and II of Fig.~\ref{ThreeStages}.) 
Since these up spin islands contributed in decreasing the resistance $R$, 
their abrupt disappearance has an effect of increasing the resistance $R$. 
If this positive contribution is larger than the negative one from 
the entrance of the domain wall, then the resistance jump becomes positive. 
In the experiment of Ref.~\cite{Otani}, such positive resistance 
jumps were indeed observed. 

In an experiment of a ferromagnetic wire, 
one can make an artificial trapping potential to trap a single domain wall 
at a fixed position in the wire, 
by verying locally the width or the shape of the wire. 
When the trap works well, 
a negative or positive resistance jump is always expected. 

In our explanation of the present phenomena, 
we assumed that a domain wall and an island in a magnetic 
domain structure can freely move and flexibly combine with an impurity 
to reduce the resistance. However, this assumption is not 
necessarily valid for a general ferromagnetic wire. 
In fact, a positive resistance contribution from a magnetic domain wall 
or a magnetic domain structure was observed for some materials 
\cite{positiveR} which have a rigid magnetic domain structure. 

We wish to thank Shinji Nonoyama, Yoshichika Otani, and 
Hal Tasaki for many useful comments.


\begin{references}
\bibitem{MP} 
The maximum transmission probability does not immediately 
lead to the minimum heat generation. The more precise explanation is 
as follows: Note that, when the total current through a wire is fixed, 
the total resistance is minimized 
owing to the principle of minimum heat generation. 
Following the Landauer formula for a quantum wire, 
the total conductance is proportional to 
the total transmission probability through the wire. 
This implies that the resistance is 
a decreasing function of the transmission probability. 
Thus the transmission probability (or transmission 
probabilities through local impurities potentials) 
must increase to reduce the heat generation.
\end{references}
\end{document}